# The Importance of Multiple Observation Methods to Characterize Potentially Habitable Exoplanets: Ground- and Space-Based Synergies

Giada Arney* [1,2], Natasha Batalha[3], Nicolas Cowan[4], Shawn Domagal-Goldman[2,5], Courtney Dressing[6], Yuka Fujii[7,8], Ravi Kopparapu[2,5], Andrew Lincowski[2,9,10], Eric Lopez[1], Jacob Lustig-Yaeger[2,9,10], Allison Youngblood[11]

* Corresponding author: giada.n.arney@nasa.gov, 301-614-6627

[1] Planetary Systems Laboratory, NASA Goddard Space Flight Center, Greenbelt, MD
[2] Virtual Planetary Laboratory, University of Washington, Seattle, WA
[3] Space Telescope Science Institute Baltimore, MD
[4] Department of Earth and Planetary Science, McGill University, Montreal, QC, Canada
[5] Planetary Environments Laboratory, NASA Goddard Space Flight Center, Greenbelt, MD
[6] Department of Astronomy, University of California Berkley, Berkley, CA
[7] NASA Goddard Institute for Space Studies, New York, NY
[8] Earth-Life Science Institute, Tokyo Institute of Technology, Ookayama, Meguro, Tokyo, Japan
[9] Department of Astronomy, University of Washington, Seattle, WA
[10] Astrobiology Program, University of Washington, Seattle, WA
[11] Exoplanets and Stellar Astrophysics Laboratory, NASA Goddard Space Flight Center, Greenbelt, MD

# 1. Types of potentially habitable exoplanets

The discovery of a truly habitable exoplanet would be one of the most important events in the history of science. However, the nature and distribution of habitable environments on exoplanets is currently unconstrained. The exoplanet revolution teaches us to expect surprises. Thus, versatile, capable observatories, and multiple observation techniques are needed to study the full diversity of habitable environments. Table 1 summarizes the challenges and opportunities of observing planets orbiting M dwarf vs. FGK dwarfs, which are best targeted with different methods.

|  | **Challenges** | **Opportunities** |
|---|---|---|
| **Potentially habitable planets orbiting M dwarfs** | Potentially significant barriers to habitability (e.g. desiccation and atmospheric loss from the super-luminous pre-main sequence phase; stellar activity-driven atmospheric loss) | Most common types of stars (~75%) <br><br> Planets with significantly different histories to Earth valuable for comparative planetology <br><br> Transits can be observed with JWST/ground-based telescopes (shorter orbital periods, deeper transit depths) |
|  | Difficult to observe via direct imaging due to the small planet-star angular separation for all but the closest targets | Moderate planet-star contrast ratio ($10^{-8}$ in reflected light) |
| **Potentially habitable planets orbiting FGK dwarfs** | Less common types of stars <br><br> Difficult to observe in transit (long orbital periods, small transit depths, less probable transits) | Less significant barriers to habitability: moderate stellar evolution and activity <br><br> Observable with potential future space-based telescopes in direct imaging due to larger planet-star angular separation |
|  | Challenging planet-star contrast ratio ($10^{-10}$ for a Sun-Earth twin in reflected light) | Understand Earth in the context of planets with similar histories. Is Earth typical? |

Table 1. Challenges and opportunities for HZ planets around M dwarfs and FGK dwarfs.

## 1.1   Potentially Habitable Planets Around M Dwarfs

M dwarfs comprise roughly 75% of the stars in our galaxy (Henry et al. 2006), and early M dwarfs may harbor habitable zone (HZ) planets at a relatively high rate of roughly 0.25 per early M dwarf (Dressing & Charbonneau 2015). Planets in the HZs of M dwarfs may experience significantly different processes compared to planets orbiting FGK stars. During formation, M dwarfs undergo an extended super-luminous (1-2 orders of magnitude brighter) pre-main sequence (PMS) phase that can last for to up to $10^9$ years. Depending on initial water inventory, this can cause extreme early water loss, such that HZ planets around M dwarfs may instead be desiccated Venus-like worlds (Luger & Barnes 2015; Ramirez & Kaltenegger 2014). However, the damage wrought by the PMS phase might be avoided by late volatile delivery (Morbidelli et al. 2000; Wang & Becker 2013), or late migration into the HZ by volatile-rich planets. The resonant orbits of the TRAPPIST-1 planets suggests that radial migration occurred (Luger et al. 2017), but this migration timescale was likely much shorter than the star's PMS phase (Bolmont et al. 2016). In addition, M dwarfs tend to be highly active (West et al. 2015). Their high x-ray luminosities (Shkolnik et al. 2015) and frequent energetic flaring (e.g. MacGregor et al. 2018) may lead to severe – perhaps total – atmospheric loss on rocky HZ planets (e.g. Owen & Mohanty 2016; Airapetian et al. 2017; Garcia-Sage et al. 2017).

Although planets orbiting M dwarfs face barriers to habitability, they are invaluable from a comparative planetology standpoint, and may even reveal novel planetary processes. For instance, extreme tides (Barnes et al. 2013) can produce synchronous rotation and qualitatively different circulation patterns than we experience on Earth (Fujii et al. 2017; Kopparapu et al. 2017; Yang et al. 2013). If we find that some M dwarf HZ planets are truly habitable, we will learn that habitability can be tenacious even in the face of significant challenges, which has important implications for the distribution and frequency of habitable worlds and the emergence of life in the galaxy.

### 1.2 Potentially Habitable Planets around Sun-like Stars

Observations of planets orbiting Sun-like stars should be sought as a high priority, as they do not face the multiple barriers to habitability posed by M dwarfs from extreme activity and the extended PMS phase. Studies of planets around Sun-like stars will place our planet into a larger cosmic context and allow us to understand how typical our world is. While the challenges of observing planets around Sun-like stars are not trivial, if M dwarfs are truly inhospitable hosts, planets around FGK stars represent our best chance of finding life elsewhere. New facilities beyond those currently planned may be required for these observations.

### 2. Characterization of potentially habitable exoplanets

Different observational methods are needed to observe different types of potentially habitable exoplanets (Table 2). By using multiple methods, we can also increase our confidence of correctly interpreting the environmental conditions and potential for life on exoplanets (e.g. habitable surface temperatures inferred from the thermal IR could be corroborated by detection of liquid surface water in reflected visible/NIR light). Additionally, by observing multiple types of planets (e.g. planets orbiting M dwarf vs Sun-like stars), we can compare their prospects for habitability in the context of diverse astrophysical environments.

Planets orbiting M dwarfs are most accessible using transit techniques due to their short orbital periods and large transit depths. Importantly, if observed via transit and radial velocity (RV), both mass and radius can be known (transit timing variations offer an additional method to determine masses in multi-planet systems; Agol & Fabrycky 2017).

Due to the geometric transit probability alone, the majority of potentially habitable planets will be invisible to transit surveys independent of observatory noise floors or instrumental capabilities. This applies to all stars, but particularly to HZ planets orbiting Sun-like stars due to infrequent and shallow transits, and low transit likelihood. Thus, HZ planets around-Sun like stars are best investigated with direct imaging. Unlike transits, direct imaging can probe surface environments, providing crucial information on near-surface conditions and the existence of liquid surface water. Coronagraphy has already advanced through ground-based developments (Lagrange et al. 2010; Marois et al. 2008), while starshades are an alternative method for starlight suppression (Cash 2006). If WFIRST included a coronagraph or starshade, it would greatly enhance these technologies.

Below, we summarize the capabilities offered by the James Webb Space Telescope (JWST), future large ground-based observatories, and potential future space telescopes including the Origins Space Telescope (OST), the Habitable Exoplanet Imaging Mission (HabEx), and the Large Ultraviolet Optical Infrared Surveyor (LUVOIR) for future observations of HZ exoplanets.

|  | **Planets around M dwarfs (transiting)** | **Planets around M dwarfs (non-transiting)** | **Planets around Sun-like stars** |
|---|---|---|---|
| **JWST** | Transit transmission spectra, secondary eclipses, phase curves | Thermal phase curves | |
| **Large ground-based facilities** | RV mass measurements, transits, reflected/thermal phase curves, direct imaging for close targets | RV mass measurements, reflected/thermal phase curves, direct imaging for close targets | Thermal emission detection may be possible, RV mass measurements |
| **Potential future space-based facilities** | Thermal phase curves & secondary eclipses (OST), transit spectra (OST, HabEx, LUVOIR), direct imaging for close targets (HabEx, LUVOIR) | Thermal phase curves (OST), direct imaging for close targets (HabEx, LUVOIR) | Direct imaging (HabEx, LUVOIR) |

Table 2. Potential capabilities of future facilities for characterizing different rocky HZ exoplanets.

## 2.1 Observations with JWST

JWST will be capable of transit spectroscopy for $\lambda = 0.6\text{-}13$ μm and will provide the first test of whether rocky HZ M dwarf planets can possess atmospheres. It may be able to characterize the atmosphere of at least one rocky M dwarf HZ exoplanet after coadding several transits (Barstow & Irwin 2016; Batalha et al. 2015; Beichman et al. 2014; Cowan et al. 2015; Meadows et al. 2018; Morley et al. 2017). However, JWST may encounter a noise floor of 20 ppm (near-IR) - 50 ppm (mid-IR) (Greene et al. 2016). The true impact of this noise floor will not be known until after telescope commissioning and early observations. Fortunately, planets orbiting smaller stars like TRAPPIST-1 will produce larger transit depths, possibly with some features exceeding 100 ppm (Morley et al. 2017). In any case, careful target selection for JWST will be vital (Morgan et al. 2018).

Transit spectroscopy will not sense the planetary surface and can be complicated by the effects of refraction (Bétrémieux & Kaltenegger 2014; Misra et al. 2014), and/or clouds and hazes (e.g. Kreidberg et al. 2014), which obscure information about the lower atmosphere. However, JWST-era spectra should be less impacted by the presence of clouds past 2.5 μm (Batalha & Line 2017).

JWST will also be capable of observing secondary eclipses of transiting exoplanets orbiting small stars in the mid-IR (> 5 μm). Such measurements sense light emitted by and reflected off the dayside of exoplanets and have been used to study atmospheres of hot Jupiters (e.g. Charbonneau et al. 2005). Secondary eclipses are less impeded by clouds/hazes than transit transmission, and therefore may enable searches for tropospheric water and other condensable or surface-produced molecules. JWST may also be able to resolve the shape of thermal phase curves for temperate terrestrial planets, including non-transiting ones, in nearby M-dwarf systems (Boutle et al. 2017; Kreidberg & Loeb 2016; Meadows et al. 2018; Turbet et al. 2016). Thermal phase curves can reveal whether a planet is likely to possess an atmosphere, and show how its thermal energy is distributed.

## 2.2 Observations with Extremely Large Ground-Based Telescopes

In coming decades, extremely large (20-40 m) ground-based facilities will come online and observe planets orbiting nearby M dwarfs, complementing JWST observations. These facilities include the Giant Magellan Telescope (GMT, 24.5 m diameter), the Thirty Meter Telescope (TMT, 30 m diameter), and the European Extremely Large Telescope (ELT, 39 m diameter).

High-resolution (high-R) spectroscopy (e.g. R > 100,000) may allow characterization of planet atmospheres from the ground (e.g. Snellen et al. 2013). Ground-based telescopes are better for this technique than potential future space-based telescopes (Section 2.3) because their larger collecting areas that make them less vulnerable to detector noise limitations (Wang et al. 2017). High-R ground-based spectroscopy will complement JWST's lower R broadband spectra, enabling searches for narrow features that JWST might not resolve such as the $O_2$ A-band at 0.76 μm, an important biosignature (Rodler & López-Morales 2014).

Direct imaging of M dwarf planets (e.g. Proxima Centauri b; Lovis et al. 2017) will be helped by the better planet-star contrast ratio compared to solar-type stars ($10^{-8}$ for an M5-Earth pair; $10^{-10}$ for a Sun-Earth twin at optical wavelengths). Adaptive optics that make these coronagraphic observations possible will allow access to the NIR (Bouchez et al. 2014; Lloyd-Hart et al. 2006). However, the wavelengths accessible to coronagraphs are also limited by the inner working angle (IWA). The IWA scales with $c\lambda/D$ (D = telescope diameter, c = constant of order unity), and it denotes the smallest star-planet angle at which a planet can be observed. Planned ground-based telescopes are larger than space-based telescope concepts being studied (Section 2.3), so their smaller IWAs may access longer wavelengths and/or more distant M dwarfs. Small IWAs also more easily access planets near crescent phase, allowing searches for ocean glint, which is strongest towards crescent (Robinson et al 2011). Ground-based telescopes may even be able to observe thermal emission from HZ planets around Sun-like stars where the contrast ratio is better (Quanz et al. 2014), providing complementarity to observations with space-based telescopes described below. Additionally, ground-based telescopes are also useful for obtaining planet masses via the RV method, which is important for interpreting atmospheric spectra.

**2.3 Observations with Potential Future Space Telescopes**

NASA is studying three flagship concepts for consideration in the 2020 astrophysics decadal survey that could observe potentially habitable exoplanets. Space-based telescopes will not have to stare through the atmosphere, which contains many of the gases we are searching for and therefore must be carefully corrected for in ground-based spectroscopy. One of these missions, OST (6.5-9.1 m) is a 6-660 μm infrared observatory that could perform transit spectroscopy, phase curve, and secondary eclipse measurements of M dwarf HZ planets at greater precision than JWST. Thus, it could provide more detailed characterization of targets already observed by JWST/the ground (including confirming the presence of possible biosignatures with higher fidelity), and it could conduct new observations of targets inaccessible to JWST.

Two other flagship concepts under consideration for the 2020 astrophysics decadal survey focus heavily on direct imaging of potentially habitable exoplanets orbiting Sun-like stars at UV-optical-NIR wavelengths: HabEx (4-6.5 m) and LUVOIR (9-15 m). In contrast to transit transmission, clouds could actually aid detection of molecules in direct imaging by increasing the contrast between the continuum and bottom of absorption bands. Direct imaging can also search for surface biosignatures, surface water, and can detect surface inhomogeneity and rotation periods. HabEx will search for Earth-like planets around tens of nearby FGK stars with a coronagraph and/or starshade, while LUVOIR will search 100s of stars for Earths with a coronagraph. Because these missions would be the first to comprehensively characterize HZ exoplanets around Sun-like stars, they will access an entirely new population of worlds whose potential for habitability may be more robust than M dwarfs. However, these telescopes could directly image the closest M dwarf HZ planets, providing synergy with possible transit observations. LUVOIR and HabEx could also observe transits of the same targets accessible to JWST and ground-based telescopes, in a complementary shorter wavelength range. UV transit spectroscopy in particular could probe escaping atmospheres (Bourrier et al. 2017; Ehrenreich et al. 2015). Thermal radiation would be inaccessible to reflected light observations, but longer thermal wavelengths may be accessible from the ground (Section 2.2).

## 3. Conclusion

The exoplanet revolution has already produced major surprises (e.g. the prevalence of super Earths), and the characterization of potentially habitable exoplanets with future observatories will surely continue this trend. The most complete understanding of the nature and distribution of habitable environments and life in the universe demands observations of multiple diverse planets orbiting a range of stellar masses. To do this, a variety of observatory types and observational methods are needed. M dwarf worlds are abundant, but their prospects for habitability may be perilous. Even if not habitable, studies of such worlds will be invaluable for comparative planetology, and they will be important targets for JWST and ground-based telescopes. In the more distant future, direct imaging of planets orbiting Sun-like stars may become feasible with large space-based direct imaging telescopes, possibly our best chance of uncovering worlds more like our own.

**References**

Agol, E & Fabrycky, D. 2017, arXiv:1706.09849
Airapetian, et al. 2017, Astrophys J, 836
Barnes, R., et al. 2013, Astrobiology, 13, 225
Barstow, J. K., & Irwin, P. G. J. 2016, MNRAS, 5, 1
Batalha, N. E., & Line, M. R. 2017, Astron J, 153 151
Batalha, N. et al. 2015, arXiv:1507.02655
Beichman, C., et al. 2014, Astron Soc Pacific, 126, 1134
Bétrémieux, Y., & Kaltenegger, L. 2014, Astrophys J, 791
Bolmont, E., et al. 2016, MNRAS, 11, 1
Bouchez, A. H. et al. 2014, Proc. SPIE, 9158
Bourrier, V. et al. 2017, Astron Astrophys, 599, L3
Boutle, I. A. et al. 2017, Astron. Astrophys. 120, 1
Cash, W. 2006, Nature, 442, 51
Charbonneau, D. B., et al. 2005, Astrophys J, 626, 523
Cowan, N. B., et al. 2015, PASP, 311
Dressing, C. D., & Charbonneau, D. 2015, Astrophys J, 807, 1
Ehrenreich, D., et al. 2015, Nature, 522, 459
Fujii, Y., et al. 2017, Astrophys J, 848
Garcia-Sage, K., et al. 2017, Astrophys J, 844
Greene, T. P., et al. 2016, Astrophys J, 817
Henry, T. J., et al. 2006, Astron J, 132, 2360
Kopparapu, R. et al. 2017, Astrophys J, 845
Kreidberg, L., et al. 2014, Nature, 505, 69
Kreidberg, L., & Loeb, A. 2016, Astrophys J Lett, 832, 1
Lagrange, A., et al. 2010, Science, 4, 57
Lloyd-Hart, M., et al. 2006, SPIE Proc 6272
Lovis, C., et al. 2017, Astron Astrophys, 599, A16,
Luger, R., & Barnes, R. 2015, Astrobiology, 15, 119
Luger, R., et al. 2017, Nat Astron, 1, 1
MacGregor, M. et al. 2018, Astrophys J Lett, 855, L2
Marois, C., et al. 2008, Science, 322, 1348
Meadows, V. S., et al. 2018, Astrobiology, 133
Misra, A., Meadows, V., & Crisp, D. 2014, Astrophys J, 792, 61
Morbidelli, et al. 2000, Meteorit Planet Sci, 35, 1309
Morgan, J., et al. 2018, 14, 1
Morley, C. V., et al. 2017, Astrophys J, 850, 121
Owen, J. E. & Mohanty, S., 2016, MNRAS, 459, 4088
Quanz, S. et al. 2015. Int. J of AsBio, 14, 2.
Ramirez, R. M., & Kaltenegger, L. 2014, Astrophys J Lett, 797
Rodler, F., & López-Morales, M. 2014, Astrophys J, 781
Robinson et al. 2011. ApJL., 721
Snellen, I. A. G., et al. 2013, Astrophys J, 764
Shkolnik, E. L., Barman, T. S., 2014, Astron J, 148, 64
Turbet, et al. 2016, Astron Astrophys, 596
Wang, J., et al. 2017, Astron J, 153, 183
Wang, Z., & Becker, H. 2013, Nature, 499, 328
West, A. A., et al. 2015, Astrophys J, 812
Yang, J., et al. 2013, Astrophys J Lett, 771